\def\gta{\ifmmode {\mathbin{\lower 3pt\hbox   %> or of order
    {$\,\rlap{\raise 5pt\hbox{$\char'076$}}\mathchar"7218\,$}}}
    \else {${\mathbin{\lower 3pt\hbox
    {$\rlap{\raise 5pt\hbox{$\char'076$}}\mathchar"7218\,$}}}
    $}\fi}
\def\lta{\ifmmode {\,\mathbin{\lower 3pt\hbox   %< or of order
    {$\,\rlap{\raise 5pt\hbox{$\char'074$}}\mathchar"7218\,$}}}
    \else {${\mathbin{\lower 3pt\hbox
    {$\rlap{\raise 5pt\hbox{$\char'074$}}\mathchar"7218\,$}}}
    $}\fi}

\def\gapprox{\gta}
\def\lapprox{\lta}

\documentclass{ws-procs975x65}

\begin{document}

\title{Observational Evidence for Intermediate-Mass Black Holes
in Ultra-Luminous X-ray Sources}

\author{
E.~J.~M. COLBERT%
\footnote{%
\uppercase{P}resent address:
\uppercase{C}atholic \uppercase{U}niversity of \uppercase{A}merica, 
\uppercase{D}ept. of \uppercase{P}hysics, 
620 \uppercase{M}ichigan \uppercase{A}ve. \uppercase{NE}, 
\uppercase{W}ashington, \uppercase{DC}~~20064, \uppercase{USA}  
}
}

\address{Johns Hopkins University, \\
Center for Astrophysical Sciences, \\
Department of Physics and Astronomy, \\
3400 North Charles Street, \\
Baltimore, MD~~21218, USA\\ 
E-mail: ed@colbertastro.org\\
}

\author{M.~.C. MILLER}

\address{Univeristy of Maryland, \\
Department of Astronomy, \\
College Park, MD~~20742, USA\\
E-mail: miller@astro.umd.edu\\
}

\maketitle

\abstracts{
Evidence is mounting that some Ultra-luminous X-ray sources (ULXs)
may contain accreting intermediate-mass black holes (IMBHs).
We review the current observational evidence for IMBH-ULXs.
While low-luminosity ULXs with
L$_X \lapprox$ 10$^{39.5}$ erg~s$^{-1}$ (assuming isotropic emission)
are consistent with mildly X-ray
beamed high-mass X-ray binaries, there are a considerable number of ULXs
with larger X-ray luminosities that are not easily explained by these
models.  Recent 
high-S/N XMM X-ray 
spectra are showing an increasing number of ULXs with 
``cool disks'' -- accretion disks with multi-color blackbody inner 
disk temperatures 
kT$_{in} \sim$ 0.1$-$0.2 keV, consistent with accreting IMBHs.  
Optical emission-line studies of
ULX nebulae provide useful measurements of X-ray energetics, and can thus
determine if the X-rays are emitted isotropically.
Analysis of an optical spectrum of the Ho~II ULX nebulae implies
an X-ray energy source with $\sim$10$^{40}$ erg~s$^{-1}$ is present,
suggesting an isotropically-emitting IMBH.
The 
spatial coincidence of ULXs with dense star clusters (young clusters and
globular clusters) suggests that IMBHS formed in these clusters could be
the
compact objects in the associated ULXs.  Quasi-periodic oscillations and
frequency breaks in XMM power-density spectra of ULXs also 
suggest that the black hole masses are more consistent with IMBHs than 
stellar-mass black holes.  Since {\it all of these ULXs with evidence for 
IMBHs are high-luminosity ULXs, i.e., L$_X \gapprox$ 10$^{40}$ erg~s$^{-1},$} 
we suggest that this
class of ULXs is generally powered by accreting IMBHs.
}

\section{Introduction}

It has long been suspected that black holes of masses  $\sim
10^2-10^4\,M_\odot$ may form in, for example, the centers of dense
stellar clusters (e.g., Wyller 1970; Bahcall \& Ostriker 1975; Frank \&
Rees 1976; Lightman \& Shapiro 1977; Marchant \& Shapiro 1980; Quinlan
\& Shapiro 1987; Portegies Zwart et al. 1999; Ebisuzaki et al. 2001).
However, for many years there was no observational evidence for such a mass
range.  In roughly the last decade, X-ray and optical
observations have revived this possibility.  If such black holes exist,
especially in dense stellar clusters, they have a host of implications,
especially for cluster dynamical evolution and the generation of
gravitational waves.

In this article, we
discuss the evidence for intermediate-mass black
holes (IMBHs) in Ultra-Luminous X-ray sources (ULXs).  ULXs are extra-nuclear
point sources that have
X-ray
fluxes many times the angle-averaged flux of a $M \gapprox 20\,M_\odot$
black hole accreting at the Eddington limit.  
Evidence for IMBHs in globular clusters and other astrophysical objects has
been discussed in several recent review articles, such as 
van~der~Marel (2003),
and Miller \& Colbert (2004).
We focus here on the ULXs, which we regard as having the most
convincing evidence for IMBHs.

\section{A Brief History of ULXs}

If we consider stellar-mass black holes (BHs) to have a maximum mass of
$\approx$20~M$_\odot$ (e.g. Fryer \& Kalogera 2001), then the  Eddington
luminosity (the limiting bolometric luminosity; see Eqn. 2) 
of an  ``intermediate-mass'' BH is $\gapprox$ 3 $\times$
10$^{39}$ erg~s$^{-1}$. 
The limiting X-ray  luminosity in the 2$-$10 keV band, for example, will
be a factor of a  few$-$10 times smaller, and it will be dependent on
the metallicity as well. The lower limit to the X-ray luminosity for a
ULX is defined to be 10$^{39.0}$ erg~s$^{-1}$.  In practice, this limit
distinguishes the ``normal'' BH XRBs  ($L_X \lapprox$10$^{39.0}$
erg~s$^{-1}$) found in our Galaxy from the intriguingly more luminous
ULXs found in some nearby galaxies.  The upper  limit for L$_X$ for ULXs
is not specified, but  usually objects have L$_X <$ 10$^{40.5}$
erg~s$^{-1}$, and most of them have L$_X <$ 10$^{40.0}$ erg~s$^{-1}$.
Quasars, supernovae, and other galaxy nuclei are usually omitted,
although some workers (e.g. Roberts et al. 2002a) include X-ray luminous
supernovae.

ULXs were named as such by several Japanese
workers who analyzed spectra from the Japanese X-ray satellite ASCA
(Mizuno et al. 1999, Makishima et al. 2000).
Here,
``ultra-luminous'' is gauged with respect to ``normal'' X-ray  binaries.
Another term that is used is ``Intermediate-luminosity X-ray Objects,''
(IXOs), which simply indicates that their X-ray luminosities are
intermediate between those of ``normal''  stellar-mass BH~XRBs, and AGNs.

ULXs were observed as early as the 1980s, when  extensive X-ray
observations of external galaxies were first performed with the Einstein
satellite.  
Many of the nearby AGNs in Seyfert galaxies were
expected to have very luminous X-ray nuclei, but it was a surprise to find
that many ``normal'' spiral galaxies also had central X-ray sources (see
Fabbiano 1989 for a review).   These X-ray sources had X-ray luminosities
$\gapprox$ 10$^{39}$ erg~s$^{-1}$, well above the Eddington value for a
single neutron star or a stellar-mass black hole.  The spatial resolution
of the most widely used instrument on Einstein (the Imaging Proportional
Counter, or the IPC, FWHM $\sim$1$^{\prime}$) is $\gapprox$1 kpc for
typical galaxy distances $\gapprox$4 Mpc, so it was not clear whether these
sources were single or multiple objects, or whether they were really
coincident with the nuclei.  
Some  possibilities included a single
supermassive black hole with a low accretion rate, a black hole with a
normal accretion rate and super-stellar mass, hot gas from a nuclear
starburst, groups of $\gapprox$10 ``normal'' X-ray binaries (e.g. Fabbiano
\& Trinchieri 1987),  or very luminous X-ray supernovae (see Schlegel
1995). The supermassive black hole scenario was not very well supported
since there was not typically any other evidence for an AGN from
observations at optical and other wavelengths.  Another possibility was
that the errors in the galaxy distances were producing artificially large
X-ray luminosities. The nearest of these interesting X-ray objects is
located in the center of the Local Group spiral galaxy M33 (see Long et al.
1981). Several other Einstein observations of similar objects are reported
in Fabbiano \& Trinchieri (1987), and a summary of Einstein observations
are given in the review article by Fabbiano (1989).  Unfortunately, the
X-ray spectral and imaging capabilities of the IPC instrument were not
generally good enough to distinguish between the possible scenarios. Even
so, it was certainly realized that these very luminous X-ray sources were
not uncommon in normal galaxies, and that they certainly deserved further
attention.

The ROSAT satellite was launched into orbit in  1990, and began
producing X-ray images at $\sim$10$-$20$^{\prime\prime}$ resolution. The
highest resolution instrument was the High Resolution Imager  (HRI; PSF
$\approx$ 10$^{\prime\prime}$). The sensitivity and spatial resolution
were a significant improvement over the Einstein IPC, many more ULXs were
discovered, and several surveys were done. It was soon found that some of
the luminous Einstein sources were not coincident with the galaxy nucleus.
For example, after registering the ROSAT image
of the nearby spiral galaxy NGC~1313
with the X-ray bright supernova~1978K, Colbert et al. (1995) found
that the central Einstein source (X-1) was actually located $\sim$1$^{\prime}$
($\sim$1 kpc) NE of the center of the nuclear bar. Some 
Einstein X-ray sources
in other galaxies, are, however, still  consistent with being located
in the galaxy nucleus.  For example,  even with Chandra  accuracy
(1$^{\prime\prime}$), the M33 ULX is still coincident with the nucleus
of the galaxy, although is not thought to be an AGN, since the dynamic
mass at that position is too small and the X-ray and optical properties
are more consistent with it being an XRB-like object (Gebhardt et al.
2001, Long et al. 2002, Dubus \& Rutledge 2002).

\begin{figure}[h!]
\hskip 0.6truein
\psfig{file=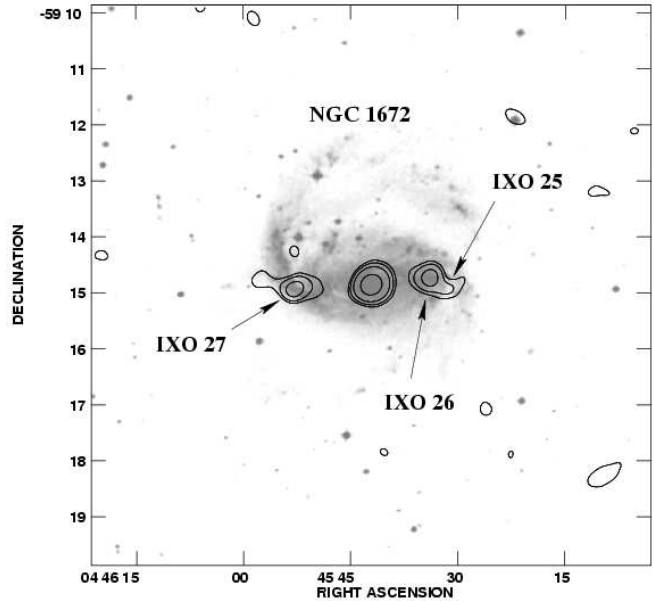,height=4.0truein}
\caption{A sample spiral galaxy with ULXs, from Colbert \& Ptak (2002).
ROSAT HRI X-ray contours from the galaxy NGC~1672 are overlaid on a B-band
DSS image of the galaxy, showing two ULXs straddling the nuclear X-ray source.
}
\label{colbertfig5spiral}
\end{figure}
The PSPC spectrometer on ROSAT had much better spectral resolution than
the Einstein IPC, but it only covered soft X-ray energies (0.2$-$2.4 keV),
and so  was of limited use for diagnosing ULX emission models.  However,
much progress was made from surveys done with the ROSAT HRI. Four large
HRI surveys of nearby galaxies  (Colbert \& Mushotzky 1999, Roberts \&
Warwick 2000,  Lira, Lawrence \& Johnson 2000, and Colbert \& Ptak 2002)
showed that off-nuclear luminous X-ray sources were actually quite common
-- present in up to half of the galaxies sampled. 
A sample spiral galaxy with two ULXs is shown 
in Figure 1 (from Colbert \& Ptak 2002).  
At the time of the first
three surveys, ULXs were not a well defined class  of objects.  
We now know that ULXs are fairly common in galaxies.  Using all of the public
HRI images for nearby galaxies, 
Ptak \& Colbert (2004) estimate that ULXs 
with L$_X$(2$-$10~keV)~$\ge$ 10$^{39}$ erg~s$^{-1}$
are present at the rate of one in every five
galaxies, on average.  When ROSAT survey work started showing that ULXs,
and thus possibly IMBHs, were quite  common, ULXs and IMBHs became a
popular topic of study.

Chandra observations of spiral and starburst galaxies have revealed
that many 
strong starburst systems have large numbers of ULXs (
$\gapprox$5 per galaxy, e.g., the Antennae,
Fabbiano, Zezas \& Murray 2001; and NGC 4485/90, Roberts et al. 2002b).
The starburst$-$ULX connection
also appears to hold at cosmologically interesting 
($z \gapprox$ 0.1) look-back times (Hornschemeier et al. 2004).  Since the
total point-source X-ray luminosity of spiral and starburst galaxies
correlates very well with the 
star-formation rate (SFR), and the most luminous point sources dominate this
luminosity (e.g., Colbert et al. 2004), many workers have argued that ULXs
are merely high-mass X-ray binaries (HMXBs) created in the starburst
(e.g., Grimm et al. 2003), and they emit X-rays anisotropically 
by mild X-ray beaming (e.g., King et al. 2001, King 2003), and thus with
a lower X-ray luminosity than if isotropic.
The low-luminosity ULXs (L$_X \lapprox$ 5 $\times$ 10$^{39}$ erg~s$^{-1}$) 
are consistent with this mild-beaming HMXB model,
and these sources dominate the population
of all ULXs (L$_X \ge$ 10$^{39.0}$ erg~s$^{-1}$; see Fig 2., and 
Ptak \& Colbert 2004).  Thus, beaming models could explain the 
implied 
``artificially high" X-ray luminosities for many ULXs.  
However, it is not know how IMBHs form, or how IMBH-ULXs ``turn on'', 
so their numbers could also be correlated with the SFR.
ULX observational diagnostics (X-ray or otherwise) have not yet 
progressed enough to be certain whether ULXs associated with
star formation (a) have IMBHs or stellar-mass BHs, (b) are beamed or
isotropic, or (c) are fed by high-mass stellar companions, or by another
source.

\begin{figure}[h!]
\hskip 2.0truein
\psfig{file=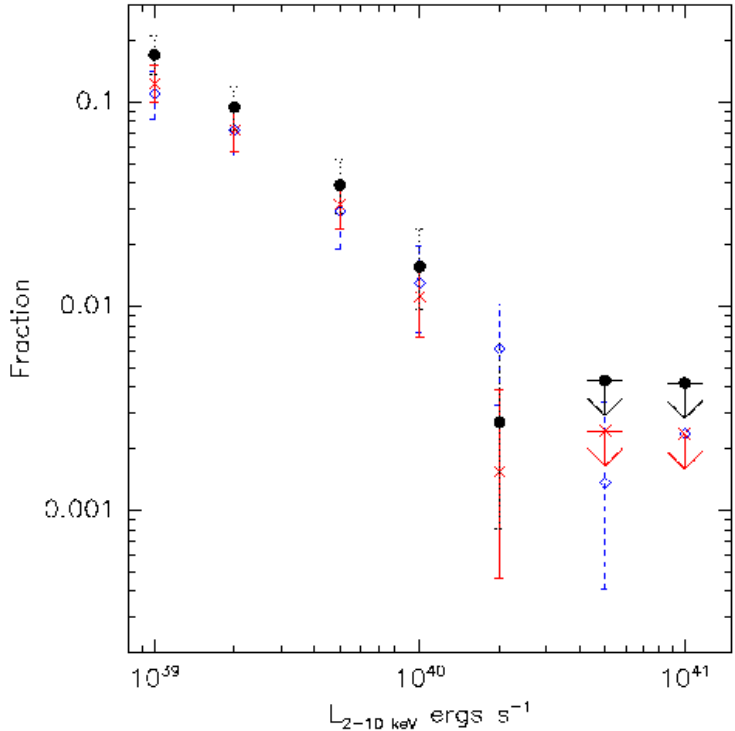,height=4.0in}
\caption{
The frequency of ULX occurrence in galaxies 
(i.e., expected number per galaxy)
as a function of L$_X$, from
Ptak \& Colbert (2004).
Diamond symbols denote statistics for for sources within a 
radius of D$_{25}$, while crosses are for sources with $r <$ 0.5 D$_{25}$.
Solid circular symbols are for spiral galaxies only ($r <$ 0.5 D$_{25}$).
}
\label{ptakcolbertfig3}
\end{figure}

\section{Observational Evidence for IMBHs in ULXs}

In the following six subsections, we describe the observational results
that
support the existence of IMBHs in some ULXs.   We emphasize that we are
not promoting that {\it all} ULXs are accreting IMBHs, but merely 
showing that some of them show very good evidence for having them.

\subsection{Extreme X-ray Luminosities}

Since one usually has no information about the flux radiation pattern
$f_X(\Omega)$ emitted by the X-ray source, it is common to assume
an ``isotropic'' X-ray
luminosity
L$_X$, as if the radiation pattern is uniform in all directions:
\begin{equation}
L_X = \int
\negthinspace
\negthinspace
\negthinspace
\negthinspace
\int d\Omega R^2 f_X(\Omega) = 4 \pi R^2 F_X,
\end{equation}
where F$_X$ is the observed X-ray flux, $R$ is the distance to the
source, and $f_X(\Omega)$ is the flux emitted per unit solid angle
in a particular direction.
For accretion around a black hole, in which the matter is
highly ionized and electron scattering is the most important form of
opacity, a source of mass $M$ that accretes and radiates {\it
isotropically} therefore cannot have an isotropic luminosity that exceeds
the Eddington luminosity
\begin{equation}
L_E=
{{4\pi GMm_p}\over{\sigma_T}}=1.3\times 10^{38}\left({{M}\over{M_\odot}}
\right)\,{\rm erg\ s}^{-1}\; ,
\end{equation}
assuming cosmic composition.  Here, 
$\sigma_T=6.65\times 10^{-25}$~cm$^2$ is the Thomson scattering
cross section.  
If the accretion or radiation is anisotropic, there is no
fundamental reason why the luminosity cannot exceed $L_E$ by an arbitrary
factor (e.g., King et al. 2001, Begelman 2002).
Beaming of the radiation can produce a flux $f_X(\Omega)$ in a
particular direction that is much greater than the average over all
angles. However, assuming isotropy holds,
a given luminosity places a lower limit on the mass of an
accreting black hole.

IMBHs with M $\gapprox$20 M$_\odot$ have L$_E \gapprox$
3 $\times$ 10$^{39}$ erg~s$^{-1}$. In the Colbert \& Ptak (2002) ULX
catalog,
45 of 87 objects have 2$-$10 keV X-ray luminosities  L$_{X} >$ $3\times
10^{39}$~erg~s$^{-1}$. Eleven objects have L$_{X} >$
$10^{40}$~erg~s$^{-1}$, which corresponds to quasi-isotropic sources with
masses $M>70\,M_\odot$. 
Based on the statistical results from Ptak \& Colbert (2004), these
``high-luminosity'' ULXs are $\gapprox$5$-$10\% of the total ULX population.
Thus, one or more
high-luminosity (L$_X \gapprox$ 10$^{40}$ erg~s$^{-1}$)
objects are expected in every 
$\sim$100 galaxies (see Figure 2), or $\gapprox$10$^7$ in the Universe,
assuming $\gapprox$10$^{10}$ galaxies.
Thus, the potential for IMBHs is clearly present.
Some galaxies, such as NGC~4038/9 (``The Antennae'') have $\gapprox$10 ULXs
with masses $\gapprox$10$-$(few)100 (by the Eddington argument), if the
X-rays are not beamed 
(see Fabbiano et al.
2001). The brightest ULX yet observed, in the galaxy M82 (e.g., see 
Ptak \& Griffiths 1999; Matsushita et al. 2000; Kaaret et al. 2001), has a
peak X-ray luminosity of $9\times 10^{40}$~erg~s$^{-1}$ (Matsumoto et al.
2001), implying a mass $M>700\,M_\odot$ by Eddington arguments.   This is
well beyond what is expected from stellar evolution.
In addition,
even a star that starts its life with a high mass may lose most of it to
winds and pulsations, leaving behind a black hole of mass $M\lta
20\,M_\odot$ if it forms with roughly solar metallicity (e.g., Fryer \&
Kalogera 2001). Objects of such mass must either have accumulated most of
their matter by some form of accretion, or have formed in some other epoch
of the universe.

While the possibility remains that 
these objects are under-luminous supermassive
black holes,
the locations of
the ULXs within the galaxies rule against masses more than $\sim
10^6\,M_\odot$ in many cases (e.g., see Kaaret et al. 2001, and Miller \&
Colbert 2004 for further discussion).

As mentioned, the mild beaming model of King et al. (2001) may explain
some of the low-luminosity ULXs with assumed isotropic luminosities,
but there are still a significant number of ULXs 
with L$_X \gapprox$ 5 $\times$ 10$^{39}$ erg~s$^{-1}$ (see Fig. 2).
Several groups have compiled large catalogs of ULXs from archival Chandra
data (e.g. 
Swartz, Ghosh, \& Tennant 2003, Swartz et. al 2003, in prog., and 
Ptak \& Colbert, in prog.), finding
$\sim$200$-$300 sources.  
Thus, based on a relative fraction of $\sim$5$-$10\%
for ULXs with L$_X \ge$ 10$^{40}$ erg~s$^{-1}$ (Fig. 2), we have already 
observed $\gapprox$10$-$30 ``high-luminosity''
IMBH-ULX candidates, consistent with counts of HRI ``high-luminosity'' ULXs 
(see section 3.1; Colbert \& Ptak 2002).
If this population of high-luminosity
ULXs represents the {\it observed,} 
``active'' IMBHs, there is likely an even larger
population of ``inactive'' IMBHs
(e.g. Madau \& Rees 2001), analogous to the relationship between AGNs and
``dormant'' supermassive BHs in galaxy nuclei.

\subsection{X-ray Spectra: Hot and Cool Disk Temperatures}

Although ASCA had a poor PSF (FWHM $\sim$1$'$), it had far better
sensitivity and spectral resolution than the Einstein IPC, and had much
wider spectral coverage (0.4$-$10 keV) than the ROSAT PSPC.   Therefore,
substantial progress was made using ASCA observations of ULXs in nearby
galaxies. 
A popular disk model for ASCA ULX spectra is the multi-color disk (MCD) 
blackbody model, since it was commonly used to fit X-ray spectra of ``normal'' 
BH XRBs (Mitsuda et al. 1984, Takano et al. 1994).
ULX ASCA spectra are often modeled with a (soft) MCD component
for the disk emission, plus a (hard) power-law component, which is
presumably Comptonized disk emission (e.g. see Takano et al. 1994). As for
Galactic BH~XRBs, the power-law photon index $\Gamma$ was noticed to  be
hard ($\Gamma \approx$ 1.8) in ULX low-flux states, and soft  ($\Gamma
\approx$ 2.5) in ULX high-flux states (e.g.,  Colbert \& Mushotzky 1999,
Kubota et al. 2001).
In the MCD model, 
the inferred temperature $T_{\rm in}$ of the innermost
portion of the disk is related to the mass of the black hole:
\begin{equation}
kT_{\rm in}\approx 1.2\,{\rm keV}\left(\xi\over 0.41\right)^{1/2}
\left(\kappa\over 1.7\right)\alpha^{-1/2}\left({\dot M}\over{{\dot M}_E}
\right)^{1/4}\left({{M}\over{10\,M_\odot}}
\right)^{-{{1}\over{4}}}
\end{equation}
(e.g., Makishima et al. 2000, eq. 10).  Here $\kappa$
is the spectral hardening 
factor (T$_{color}$/T$_{eff}$, also known as the color correction factor),
$\xi$
is a factor that takes
into account that the maximum temperature occurs at a radius larger
than the radius of the innermost stable circular orbit, and $\alpha=
R_{\rm in}/(6GM/c^2)$ is unity for a Schwarzschild spacetime and
$\alpha=1/6$ for prograde orbits in a maximal Kerr spacetime.
Fiducial values for $\kappa$ and $\xi$ are 1.7 and 0.41, respectively.
Thus, if $T_{\rm in}$ inferred from MCD fits is representative, one expects
lower disk temperatures (kT$_{in} \lapprox$ 0.1$-$0.2~keV)
from accreting IMBH, compared with accreting stellar-mass
black holes. 
Some detailed aspects of the application of the MCD model to ULX spectra
are given in Makishima et al. (2000).

The implications of the MCD model were problematic
for ASCA spectra. 
While very large X-ray
luminosities of $\sim$10$^{39-40}$ erg~s$^{-1}$ are explained
well
by an IMBH with sub-Eddington  accretion, the
temperature $kT_{in}$ (and radius $r_{in}$) of the inner accretion disk,
derived from MCD spectral fitting, are too  high (low) for IMBHs (Mizuno
et al. 1999; Colbert \& Mushotzky 1999; Makishima et al. 2000; Mizuno,
Kubota, \& Makishima 2001). BH XRBs with
stellar-mass black holes in our Galaxy typically have temperatures
$kT_{\rm in}\approx 0.4-1$~keV, while ULXs fit with ASCA have  
$kT_{in} \approx$ 1.1$-$1.8 keV, 
which is more consistent with LMXB micro-quasars in our
Galaxy (e.g. Makishima et al. 2000). 
In any case, many ASCA model disk temperatures were well above 
the expected $\sim$~0.1 keV.

As mentioned, one solution to this
high-temperature problem is to suppose that the
compact object is a stellar-mass BH in a HMXB and the X-ray emission
is mildly beamed so that the radiation pattern $f_X$($\Omega$) only
illuminates $\sim$10\% of the sky (King et al. 2001).
This interpretation has become quite popular, since it naturally explains
the ULX-starburst relationship (e.g., Grimm et al. 2003).

Mizuno et al. (1999),  Makishima et al. (2000), and Ebisawa et al. (2001)
offer several other
potential explanations for the ``high temperature'' problem.
For example, it is possible that the BH is a Kerr IMBH, and the resulting
frame dragging can shrink the inner radius of the accretion disk up to
$\approx$6 times less than that of a Schwarzschild BH, for which r$_{in}
\gapprox$ 3 R$_{s}$.  Therefore, $r_{in}$ can be smaller, and $T_{in}$ is
larger, as implied by the MCD models.  Kerr models work  well for ULXs as
IMBHs (e.g., Mizuno et al. 2001), but imply very high disk inclination angles
(i $\gapprox$ 80$^{\circ}$, Ebisawa et al. 2001, Ebisawa et al. 2003).

One may also relax the assumptions of the
``thin disk'' model.  For example, increasing $\kappa$, the ratio of the 
color temperature to the effective temperature, will yield higher masses
(e.g. Shrader \& Titarchik 1999), and so will increasing the correction
factor $\xi$, which adjusts for the fact that $T_{in}$ occurs at a slightly
higher radius than $r_{in}$ (see Kubota et al. 1998).  The mass $M$ is 
proportional to the product $\kappa^2\xi$.  Makishima et al. (2000) shows
that $\kappa^2\xi$ has to differ largely from values for ``normal'' BH XRBs
for the ``high-temperature'' problem to be solved.

Finally, one may completely abandon the physically thin accretion disk
model. Abramowicz et al. (1988) and Watarai et al. (2000) show that very
high accretion rates  $\dot{M} \gapprox$ 10 L$_E/c^2$ lead to an ADAF
(Advection-Dominated Accretion Flow) solution (the so-called  ``slim
disk'' model), and that this can explain the  $r_{in} \propto T_{in}^{-1}$
relationship, found for MCD fits to ASCA spectra of ULXs (Mizuno et al.
2001).  The slim-disk model allows masses to be slightly larger
($\lapprox$10$-$30 M$_\odot$), but not as large as $\sim$100 M$_\odot$.

Much effort has gone into trying to explain why the MCD temperatures
$kT_{in}$ are so high for ULXs.  However, it is possible that the ULXs are
{\it not} well represented by a simple MCD disk model after all. For
example, when simulated spectra of accretion disks are fit with  MCD
models, $r_{in}$ and/or the disk accretion luminosity are very poorly
estimated (e.g. Merloni et al. 2000, Hubeny et al. 2001). 
When the popular, but primitive MCD$+$power-law model is replaced by the
more elaborate bulk-motion Comptonization model developed by L. Titarchuk,
IMBH masses are predicted.  Shrader \& Titarchuk (2003) have further
developed this model, and use $\kappa =$ 2.6 to show its usefulness for
estimating black hole masses.

In addition,
since the PSF of ASCA is so large, ASCA spectra can be contaminated by
diffuse  X-ray emission and by X-ray emission from other point sources
positioned  extraction regions, so that a single MCD model is
inappropriate.    It was not known if this effect was significant for
ULX ASCA spectra.

XMM and Chandra observations have the advantage that their spatial
resolution ($\approx$1$^{\prime\prime}$ for Chandra,
$\approx$4$^{\prime\prime}$ for the MOS2 camera on XMM) is good enough
that contamination from other X-ray sources is not as problematic as it
is for ASCA. They also have significantly better throughput than ASCA,
which improves the signal to noise.
For XMM, the bandwidth extends up to $\sim$12 keV, a few more keV than ASCA.
This allows much better leverage for discriminating curved 
blackbody spectra from straight power-law spectra.
While ASCA ULX spectral models often required both MCD and power-law
components, Chandra spectra are often fit well with a single component
(either MCD or power-law).  In many cases, this is due to the poor quality
spectra, and especially to poor spectral coverage at energies $\gapprox$5 keV,
due to either poor sensitivity, or CCD photon pile-up problems.

\begin{figure}[h!]
\hskip 1.0truein
\psfig{file=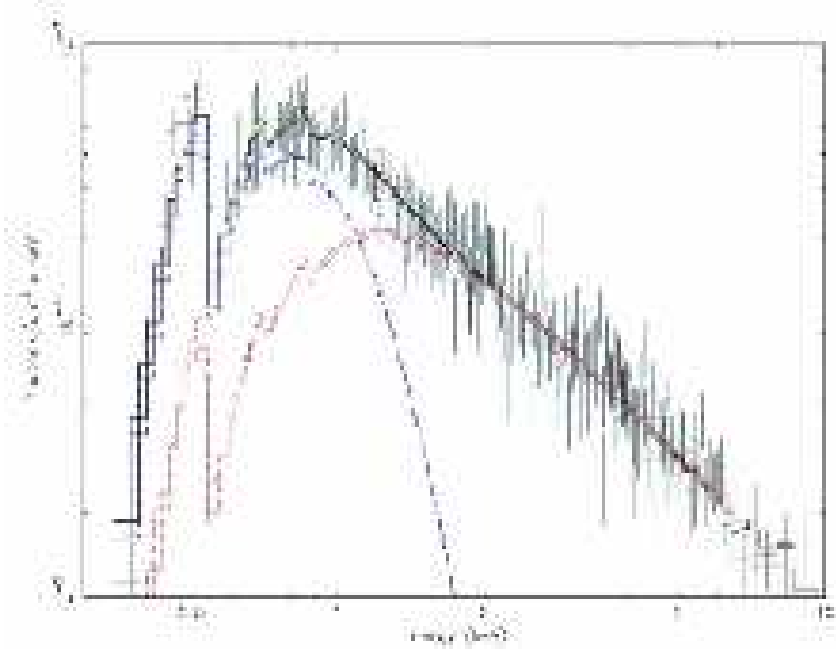,height=3.5truein}
\caption{Unfolded XMM MOS spectrum
of the ULX NGC~1313 X-1, 
from Miller et al. (2003a).
The model components (cool MCD model and a power-law) are also shown.
Absorption
of soft X-rays below $\sim$1 keV is also modeled in the fit.  
}
\label{colbertfig7}
\end{figure}
An interesting recent result from the much improved 
XMM spectra of ULXs is that, in some
cases if an MCD component is required, its inferred  temperature is 
$\sim$~0.1 keV, consistent with an accreting IMBH.
Some ULXs with ``cool disks'' were even found with ASCA, but XMM is 
revealing many more.  XMM data are able to show the significance of 
the cool MCD model component much easier than ASCA data.
For example, Miller et al.
(2003a) analyzed XMM data of the ULX NGC~1313 X-1, and found that
a two-component fit is necessary (see Figure~\ref{colbertfig7}), with an
inferred inner disk temperature $kT_{\rm in}=0.15$~keV.  In comparison,
Colbert \& Mushotzky (1999) analyzed two ASCA observations of NGC~1313 X-1;
one had a hard spectrum, with an MCD best fit temperature of  $kT=1.5$~keV,
while the other was softer and more consistent with the recent XMM analysis.
The ULX NGC~5408 X-1 is best fit with a MCD temperature  kT$_{in}
\approx$ 0.1 keV (Colbert \& Mushotzky 1999), and this is confirmed with
Chandra (Kaaret et al. 2003).  Similarly, the joint ROSAT$+$ASCA fit of the
X-ray spectrum of the ULX in Ho~II yields $kT_{in} \approx$ 0.17 keV
(Miyaji et al. 2001). 
We list X-ray observations of 
``cool disks'' IMBH candidates in Table 1.  
{\bf It
is interesting to note that the cool disk candidates all have 
high luminosities $\gapprox$10$^{40}$ erg~s$^{-1}$, precisely the
sub-sample of ULXs that are not easily explained by the anisotropic beaming
models mentioned in section 2.
}

\begin{table}[bt]
\tbl{ULXs with Cool Disks}
{\footnotesize
\begin{tabular}{@{}lcclc@{}}
\hline
{} &{} &{} &{} &{}\\[-1.5ex]
{ULX Name} & {kT$_{in}$} & {L$_X$} & {Energy} & {Comment}\\[1ex]
{} & {(keV)} & {(10$^{39}$ erg~s$^{-1}$)} & {Range (keV)} & {}\\[1ex]
\hline
{} &{} &{} &{} &{}\\[-1.5ex]
NGC 1313 X-1      & 0.12      & $\sim$10 & 2$-$10    & ASCA: Colbert \& Mushotzky 1999 \\[1ex]
                  & 0.15      & 20       & 0.2$-$10  & XMM: Miller et al. 2003a\\[1ex]
NGC 5408 X-1      & 0.13      & $\sim$10 & 2$-$10    & ASCA: Colbert \& Mushotzky 1999 \\[1ex]
                  & 0.11      & 11       & 0.3$-$8   & Chandra: Kaaret et al. 2003 \\[1ex]
Ho II ULX         & 0.2       & $\sim$10 & 0.2$-$10  & ASCA+ROSAT: Miyaji et al. 2001 \\[1ex]
Antennae (4 ULXs) & 0.1$-$0.2 & $\approx$10$-$30 & 0.3$-$10 & XMM: Miller et al. 2003b \\[1ex]
Ho IX             & 0.2$-$0.3 & 10$-$16  & 0.3$-$10  &  XMM: Miller et al. 2003c \\[1ex]
NGC 4559 X-7      & 0.12      & $\approx$20 & 0.3$-$12 &  XMM: Cropper et al. 2004, MNRAS \\[1ex]
\hline
{} &{} &{} &{} &{}\\[-1.5ex]
\end{tabular}\label{table1} }
\vspace*{13pt}
\end{table}

\subsection{``Type II'' ULXs (?) and Globular Clusters in Elliptical Galaxies}

\begin{figure}[h!]
\hskip 0.6truein
\psfig{file=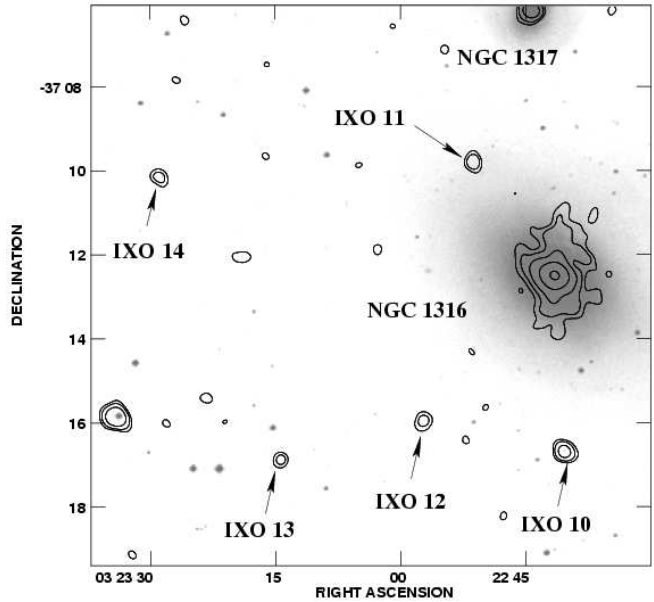,height=4.0in}
\caption{
Contours of the ROSAT HRI X-ray emission from five of the six ULXs
(IXOs) 
in the elliptical 
galaxy Fornax~A (NGC~1316), from Colbert \& Ptak (2002).
}
\label{colbertfig5ellipitical}
\end{figure}
Now that Chandra is in full operation, its combined imaging and spectral
capabilities have allowed the literature on ULXs to blossom.
The excellent imaging sensitivities of Chandra and XMM ensure
that one is likely to detect  an ULX in observations of nearby galaxies
$\gapprox$20\% of the time, for integrations of more than a few hours.
Even short ``snapshot'' observations with Chandra and  XMM
have detected a significant number of ULXs 
in both spiral and elliptical galaxies
(e.g., Sipior 2003, Foschini et al.
2002).
Most X-ray spectral modeling results
are derived for ULXs in spiral galaxies.  This is primarily due to the larger
distances of the nearby ellipticals (primarily in Virgo), and thus
the lack of available photons for spectral analysis.  If ULXs in ellipticals
are indeed a different class than those in spirals (e.g. Colbert \& Ptak 2002, 
King 2003), we 
might expect a difference in their X-ray spectral properties.  
We know that elliptical galaxies do contain ULXs (e.g., see Fig. 4), but
are these the same type of ULXs that are found in spiral and starburst
galaxies?
A census of ULXs
using all of the public ROSAT HRI data found that if one selects {\it only}
those galaxies with  detected ULXs, the elliptical galaxies with ULXs have
a larger number per galaxy than do the spiral galaxies  with ULXs (Colbert
\& Ptak 2002).   
The elliptical galaxy NGC~720 has nine ULXs, which is
nearly as many that are found in the ``Antennae'' (Jeltema et al. 2003).
Since  ellipticals are also generally more massive than spirals, it does
not imply that they are more efficient at producing ULXs, but it does imply
that the mild beaming HMXB
scenario does not work for all ULXs, since
elliptical galaxies have virtually no young stars being formed.
Thus, ULXs in elliptical galaxies are probably associated with the older
stellar population 
(population~II).

\begin{figure}[h]
\hskip 0.7truein
\psfig{file=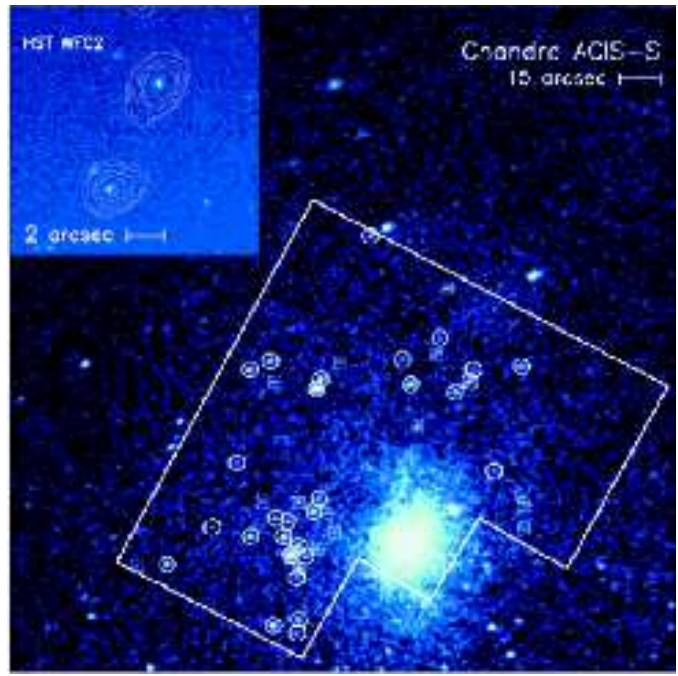,height=3.5truein}
\caption{Grey-scale representation of the smoothed Chandra ACIS-S image of 
the CD elliptical galaxy NGC~1399, from Angelini et al. (2001).  The
HST WFPC2 FOV is overlaid.  The circles show the X-ray sources 
positions that are associated with globular clusters.
}
\label{colbertfig8}
\end{figure}
The lack of confusing optical sources in elliptical galaxy halos
allows an easier
identification of
unique counterparts, compared with disk galaxies.
For example,
Angelini, Loewenstein, \& Mushotzky (2001) performed a detailed comparison
between Chandra X-ray sources in the giant elliptical galaxy  NGC~1399 and
HST counterparts, finding that 26 of the 38 sources detected at $>$3$\sigma$
were obviously associated with globular clusters (Figure~\ref{colbertfig8}).   Two of
the three ULXs are associated with globular clusters. Other groups are also finding
that there is a strong correlation between X-ray sources in elliptical
galaxies and globular clusters (e.g. Kundu, Maccarone \& Zepf 2002). 
Sarazin, Irwin, \& Bregman
(2001) find bright point sources up to $\approx 2.5\times
10^{39}$~erg~s$^{-1}$ in the elliptical galaxy NGC~4697, and conjecture
that although only 20\% of these sources are currently identified with
globular clusters, all the LMXBs may have originated in globular clusters.  In
general, low-mass X-ray binaries in early-type galaxies are strongly
correlated with globular clusters (Sarazin et al. 2003).  
It will
be exciting to learn results from follow-up optical studies  to determine the
age, metallicity and other  derivable properties for these globular clusters, and of
other globular clusters with ULXs.

These results, combined with results for ULX/young-cluster associations in
starburst systems such as the Antennae (Clark et al. 2003),
show
that there is a strong link between ULXs and star clusters, whether they
be young star-forming regions, or globular clusters, which are 100$-$1000
times older. It is of interest that there are dozens of sources in globular clusters
around NGC~1399 with L$_X >$ 10$^{38}$~erg~s$^{-1}$, given that both our
Galaxy (with $\approx$150 globular clusters, Harris 1996) and M31 (with
$\approx$300$-$400 globular clusters, Hodge 1992, Fusi Pecci et al. 1993)  
have very few
X-ray sources in globular clusters with  L$_X \gta 10^{38}$~erg~s$^{-1}$
(Hut 1993; Supper et al. 1997). Part of this may have to do with the
high number of globular clusters per unit mass around NGC~1399 (as it typical of
elliptical galaxies), which is 15 times the
average specific frequency for spiral galaxies such as the Milky Way and M31
(e.g., Kissler-Patig 1997), but there may also be evolutionary differences.

As described by Miller \& Hamilton (2002) and
Portegies~Zwart \& McMillan (2002; see review article by Miller \& Colbert 2004
for additional details), dense stellar clusters are likely sites for IMBH
formation.  There is also evidence for IMBHs in globular
clusters from HST observations 
(e.g., Gebhardt et al. 2002, Gerssen et al. 2002, van~der~Marel et al. 2002),
and the implied IMBH masses are consistent
with the extrapolation of
the tight M$_{BH}$$-$$\sigma$ relationship for galaxy bulges 
and supermassive BHs (e.g., see Gebhardt et al. 2002).

Although the globular-cluster$-$IMBH$-$``population~II~ULX'' connection
seems perfectly feasible, as mentioned, X-ray spectral diagnostics for these
``type~II'' ULXs are not good enough to determine if they are observationally
distinct from
ULXs in spiral and starburst galaxies.

Some workers have claimed that most ULXs in elliptical galaxies are 
background objects
(e.g., Irwin, Athey \& Bregman 2003; Irwin, Bregman \& Athey 2004).
However, some elliptical galaxies {\it do have} more ULXs than is expected
from background counts.
A more complete analysis of the ULXs in elliptical galaxies is sorely
needed, and should provide better insight to their nature.

\subsection{Implications from X-ray Variability Studies}

It is possible that objects other than a single accreting black hole system
could produce X-ray luminosities $\ge$10$^{39}$ erg~s$^{-1}$.  For example,
some very young ($\lapprox$ 100 yr) supernovae are known to emit
$\sim$10$^{39}$ erg~s$^{-1}$ in X-rays.  However, their X-ray
emission typically 
fades or remains constant on timescales of $\lapprox$1 yr (cf. Schlegel
1995). A cluster of $\sim$10 or more ``normal'' luminous XRBs could also
produce  $\sim$10$^{39}$ erg~s$^{-1}$.  
However,
Colbert \& Ptak (2002) estimate random variability of  $\gapprox$50\% in
over half of all ULXs, eliminating supernovae or  XRB-clusters as likely
scenarios.  The brightest X-ray source in M82 brightened by a factor of 7
between two Chandra observations three months apart (Matsushita et al.
2000).  Long-term variability of ULXs on timescales of months to years
has been noted for ULXs in many nearby spiral galaxies: M81 (Ezoe et al.
2001, La~Parola et al. 2001, Wang 2002, Liu et al. 2002), Ho~II (Miyaji
et al. 2001), M82 (Ptak \& Griffiths 1999, Matsumoto \& Tsuru 1999, Kaaret
et al. 2001, Matsumoto et al. 2001), IC~342 (Sugiho et al. 2001, Kubota et
al. 2001),  Circinus (Bauer et al. 2001), NGC~4485/90 (Roberts et al.
2002b), M101 (Mukai et al. 2002), 
and M51 (Terashima \& Wilson 2003).

Thus, variability on scales of months or longer is well-established.  For
periodic variability due to orbiting stars, Kepler's third law predicts very
short times for orbits near the BH:
\begin{equation}
P = 
1.73 \times 10^{-5} 
{ { ({a/{\rm km}})^{{3}\over{2}} }\over{ (M/M_\odot)^{{1}\over{2}} }} 
{\rm sec}
=
3.65 \times 10^{2} 
{ { ({a/{\rm AU}})^{{3}\over{2}} }\over{ (M/M_\odot)^{{1}\over{2}} }} 
{\rm days}
\end{equation}
where $a$ is the semi-major axis of the stellar orbit.  
Measurements of both the orbital period and velocity can thus put constraints
on the BH mass.
Monthly X-ray monitoring can only sample orbits around $\sim$100 M$_\odot$
BHs for stars at radial orbits of $\gapprox$1 AU (1.5 $\times$ 10$^{8}$ km),
where the probability of eclipsing is quite low.  Thus, it is important to
test for periodicity on much shorter timescales, especially when searching
for evidence for IMBHs with M $\gapprox$ 100 M$_\odot$.

There have been very few reports of variation on time scales less than a
few weeks.  There are currently three reported cases of variability on
time scales of hours, all of which have been interpreted as possibly
periodic.
Roberts \& Colbert (2003) report
aperiodic variability on timescales of a few hundred seconds from NGC 6946
X-11. Bauer et al. (2001) observed one source in the Circinus galaxy to
exhibit a count rate variation of a factor of 20, during a 67 ksec Chandra
observation. Three peaks are seen, which are consistent with a 7.5 hour
period. Bauer et al. (2001) discuss different mechanisms for this
variability, including eclipses, modulation of the accretion rate, or a
precessing jet.  Sugiho et al. (2001) observed a ULX in the spiral galaxy
IC 342 and found possible evidence for either a 31 hour or a 41 hour
period, admittedly based on only two peaks.  More recently, Liu et al.
(2002) and Terashima \& Wilson (2003) report more than 50\% variation in
count rate from a ULX in M51, with a time of 7620$\pm$500 seconds between
the two peaks seen.

It is tempting to interpret these periods as orbital periods.
This would be highly constraining for the $\sim$2 hour period of the
M51 source, and would in fact imply that the companion is a 
$\sim 0.3\,M_\odot$ dwarf (Liu et al. 2002).  However, it is
premature to draw conclusions because at this point no
source has been seen to undergo more than three cycles.  This is
a clear case in which sustained observations, especially of the
putative 2 hour period, are essential.  Only then will it be possible
to separate models in which the period is orbital (in which case it
should be highly coherent) from models in which the period is due to,
e.g., disk modes, in which the modulation could be quasi-periodic.
Furthermore, additional information (e.g. orbital velocity) is needed to
constrain the BH mass.

Variability on very short timescales (seconds to minutes) can be detected to
the same level of fractional RMS amplitude as variability on longer
timescales, but the variability of sources from XRBs to AGN tends
generally to decrease with increasing frequency.  This means that
variability at the few percent level would be detectable out to the Nyquist
frequency of observations of the brightest ULXs (e.g., to 1~Hz in the XMM
data power-density spectrum [PDS] of the M82 ULX; see
Strohmayer \& Mushotzky 2003).  It would be well
worth doing a systematic comparison of the broad-band power spectra of X-ray
binaries, ULXs, and AGN, given that one expects the maximum frequency at
which significant power exists to decrease with increasing mass.  Although
the lack of a fundamental theory of this variability limits our ability to
draw rigorous conclusions (e.g., the stellar-mass black hole LMC~X-3 has no
detected variation at $\nu>10^{-3}$~Hz; see Nowak et al. 2001), systematic
differences in the power spectra could provide insight into the nature of
ULXs.

The first, and so far only, quasi-periodic oscillation (QPO) in a ULX was
reported by Strohmayer \& Mushotzky (2003), based on XMM observations of
the brightest point source in M82.  For this high-luminosity ULX, they 
find a QPO at 54~mHz, with a
quality factor $Q\sim 5$, with a fractional rms amplitude of 8.5\%.  At the
time, the flux would imply a bolometric luminosity (if isotropic) of
$4-5\times 10^{40}$~erg~s$^{-1}$.  As discussed by Strohmayer \& Mushotzky
(2003), QPOs are usually thought to originate from disk emission,  which if
true makes this observation troublesome for a beaming interpretation.  This
is {\it not} because the frequency is low (for example, as mentioned by
Strohmayer \& Mushotzky 2003, a 67~mHz QPO has been observed with RXTE from
GRS~1915+105, which has a dynamically measured mass of $14\pm 4\,M_\odot$;
see Morgan, Remillard, \& Greiner 1997).  The problem is instead that if
the source is really a beamed stellar-mass black hole, the variability  in
the disk emission (which is nearly isotropic) would have to be of enormous
amplitude to account for the observations.  For example, even for a
$20\,M_\odot$ black hole accreting at the Eddington limit, the beaming at
$4-5\times 10^{40}$~erg~s$^{-1}$ would need to be a factor of $\sim 15$,
requiring intrinsic variability in the disk emission in excess of 100\%.
There are other sources in the XMM beam; the brightest of these sources has
an equivalent peak isotropic luminosity of $3.5\times
10^{39}$~erg~s$^{-1}$, comparable to the luminosity of $3.4\times
10^{39}$~erg~s$^{-1}$  (Strohmayer \& Mushotzky 2003).  For this source to
produce the QPO  would therefore require nearly 100\% modulation, which
seems unlikely.  
Furthermore, the QPO has a very narrow width ($\Delta\nu \approx$ 11 mHz),
which would be very difficult to sustain with the many photon paths
from reflections from the focussing X-ray mirror in the beaming scenario.
These observations therefore provide indirect evidence for
the IMBH scenario, although caution is still required because the theory of
black hole QPOs is not settled.

A recent XMM observation of the high-luminosity ULX NGC~4557 also shows 
evidence for an IMBH from X-ray timing analysis (Cropper et al. 2004).
The PDS shows a break at a frequency of 28~mHz, which suggests a BH mass
of $\sim$1300 M$_\odot$ if this break is associated with the upper
break frequency seen in PDSs of AGNs (e.g., Markowitz et al. 2003).
Note that not only does this ULX have L$_X \approx$ 2 $\times$ 10$^{40}$
erg~s$^{-1}$, and is therefore not easily explained by the mild-beaming
HMXB model, but it
also shows evidence for a cool MCD disk (Table 1).  All of
these items suggest that this ULX is an accreting IMBH.

\subsection{Energetics of X-ray Source based on Optical Spectra}

Studies of the environments around ULXs have led to some interesting
results. 
Pakull \&
Mirioni (2002) find that the ULX in the dwarf galaxy Holmberg II has an
optical nebula around it with substantial He II 4686\AA\ emission.  This
line is produced by the recombination of fully ionized helium, which
requires for its excitation a high-energy source.  
Based on models of X-ray reprocessing where the X-ray
source is located inside the nebula, Pakull \& Mirioni (2002) conclude that
the optical radiation is consistent with an isotropic, X-ray source and not
with significant X-ray beaming, which would produce far fewer 
(a factor $\lapprox$ 0.1)
EUV
ionizing
photons.  However, there is substantial uncertainty in the
correction factor from optical line flux to X-ray luminosity, so work of
this type needs to be repeated for a number of sources in order to draw
firmer conclusions.  Integral field spectroscopic observations by Roberts
et al. (2002a) indicate that many of the ULXs are actually located in
cavities free from optical line-emitting gas, although it is not clear
whether this is due to the absence of gas (e.g., gas cleared away by
shocks), or to highly ionized gas irradiated by the ULX.  Optical spectral
analyses of some of these ULX nebulae show evidence for both shocks and
photo-ionization (Pakull \& Mirioni 2003). This is intriguing, as there are
now at least two ULXs that are highly  variable (and thus are accreting
compact objects), but are directly associated with optical supernova
remnants (IC~342 X-1, Roberts et al. 2003, and MF16 in NGC~6946, Roberts \&
Colbert 2003; note that the precise mechanism for the ionization is not
rigorously established in these cases, and that jet ionization is a
possible alternative to supernova shock ionization).  Future
multiwavelength studies of optical ULX nebulae and ULXs in SNRs may provide
important clues as to how  ULXs form, or at least how they become
``active'' X-ray sources.
When emission-line spectra of individual stellar companions to ULX BHs,
assuming they exist, can eventually be
obtained, then this will help
constrain orbital the orbital velocity, and thus the BH mass (Eqn. 4).
However, high-throughput (diam. $\gapprox$ 8~m) optical telescopes will 
be required to obtain single-star spectra at distances of more than a 
few Mpc.

\subsection{Additional Evidence}

Although Fe~K lines (6.4$-$7.0 keV) are not usually strong in BH~XRBs,
Strohmayer \& Mushotzky (2003) found that the famous M82 ULX has a very 
broad Fe~K line in an XMM spectrum.  This is not easily explained by beaming
models and thus this observation provides indirect evidence for an IMBH 
in this high-luminosity ULX.
Further Fe~K line studies of ULXs will help to determine
if they can be used as reliable diagnostics for the BH mass measurements,
and ULX geometry.

\section{Summary and Conclusions}

In summary, while many of the low-luminosity ULXs with 
L$_X \lapprox$ 5 $\times$ 10$^{39}$ erg~s$^{-1}$ are consistent with 
mild-beaming HMXB models (e.g., King et al. 2001, King 2003), there are
a significant number of ULXs that are not.
The ULXs that do show evidence for isotropically-emitting
sub-Eddington IMBHs -- in the form of ``cool disks,'' powerful and narrow QPOs, 
or suggestive breaks in their PDS -- are all high-luminosity ULXs, with
L$_X \gapprox$ 10$^{40}$ erg~s$^{-1}$, precisely those that are not well
explained by mild beaming.  We emphasize that X-ray or optical/NIR
observational diagnostics
are not yet able to sytematically determine the mass, emission anisotropy,
or fuel source of ULXs.  Since the formation mechanism for IMBHs 
and IMBH-ULXs is not well understood, it is not
absolutely certain what fraction of
either the ``low-luminosity'' ULXs or the ``high-luminosity'' ULXs have 
IMBHs.  This holds for the ULXs correlated with star-formation 
in 
spiral and starburst galaxies, as well as the ``type-II'' (?) 
ULXs in elliptical galaxies.

\section*{Acknowledgments}

This work was supported in part by 
NASA grant
NAG 5-11670 at Johns Hopkins University,
and
NASA grant NAG 5-13229 
at the University of Maryland.
Portions of this review article also appear in Int. J. Mod. Phys. D 
(13, 1; 2004).

\def\reference{\bibitem}

\end{document}